\documentclass[acmlarge]{acmart}

\AtBeginDocument{%
  \providecommand\BibTeX{{%
    \normalfont B\kern-0.5em{\scshape i\kern-0.25em b}\kern-0.8em\TeX}}}

\setcopyright{acmcopyright}
\copyrightyear{2021}
\acmYear{2021}
\acmDOI{XX.XXXX/XXXXXXX.XXXXXXX}

\acmJournal{JOCCH}
\acmVolume{0}
\acmNumber{0}
\acmArticle{0}
\acmMonth{0}

\usepackage{color}
\usepackage{graphicx}
\usepackage{subfigure}

\def\ie{i.e.\ }

\def\etal{et~al.\ }

\usepackage{tabularx}
\newcolumntype{L}[1]{>{\raggedright\arraybackslash}m{#1}}
\newcolumntype{C}[1]{>{\centering\arraybackslash}m{#1}}
\newcolumntype{R}[1]{>{\raggedleft\arraybackslash}m{#1}}

\usepackage{algorithm}
\usepackage{algorithmic}

\begin{document}

\title[Towards Tangible Cultural Heritage Experiences]{Towards Tangible Cultural Heritage Experiences - Enriching VR-Based Object Inspection with Haptic Feedback}

\author{Stefan Krumpen}
\email{krumpen@cs.uni-bonn.de}
\affiliation{%
  \institution{Institute of Computer Science II, University of Bonn}
  \streetaddress{Endenicher Allee 19a}
  \city{Bonn}
  \country{GERMANY}
  \postcode{53115}
}
\author{Reinhard Klein}
\email{rk@cs.uni-bonn.de}
\affiliation{%
  \institution{Institute of Computer Science II, University of Bonn}
  \streetaddress{Endenicher Allee 19a}
  \city{Bonn}
  \country{GERMANY}
  \postcode{53115}
}
\author{Michael Weinmann}
\email{mw@cs.uni-bonn.de}
\affiliation{%
  \institution{Institute of Computer Science II, University of Bonn}
  \streetaddress{Endenicher Allee 19a}
  \city{Bonn}
  \country{GERMANY}
  \postcode{53115}
}

\renewcommand{\shortauthors}{Krumpen et al.}

\begin{abstract}
VR/AR technology is a key enabler for new ways of immersively experiencing cultural heritage artifacts based on their virtual counterparts obtained from a digitization process. In this paper, we focus on enriching VR-based object inspection by additional haptic feedback, thereby creating tangible cultural heritage experiences. For this purpose, we present an approach for interactive and collaborative VR-based object inspection and annotation. Our system supports high-quality 3D models with accurate reflectance characteristics while additionally providing haptic feedback regarding the object shape features based on a 3D printed replica. The digital object model in terms of a printable representation of the geometry as well as reflectance characteristics are stored in a compact and streamable representation on a central server, which streams the data to remotely connected users/clients. The latter can jointly perform an interactive inspection of the object in VR with additional haptic feedback through the 3D printed replica. Evaluations regarding system performance, visual quality of the considered models as well as insights from a user study indicate an improved interaction, assessment and experience of the considered objects.
\end{abstract}



\begin{teaserfigure}
	\includegraphics[width=\textwidth]{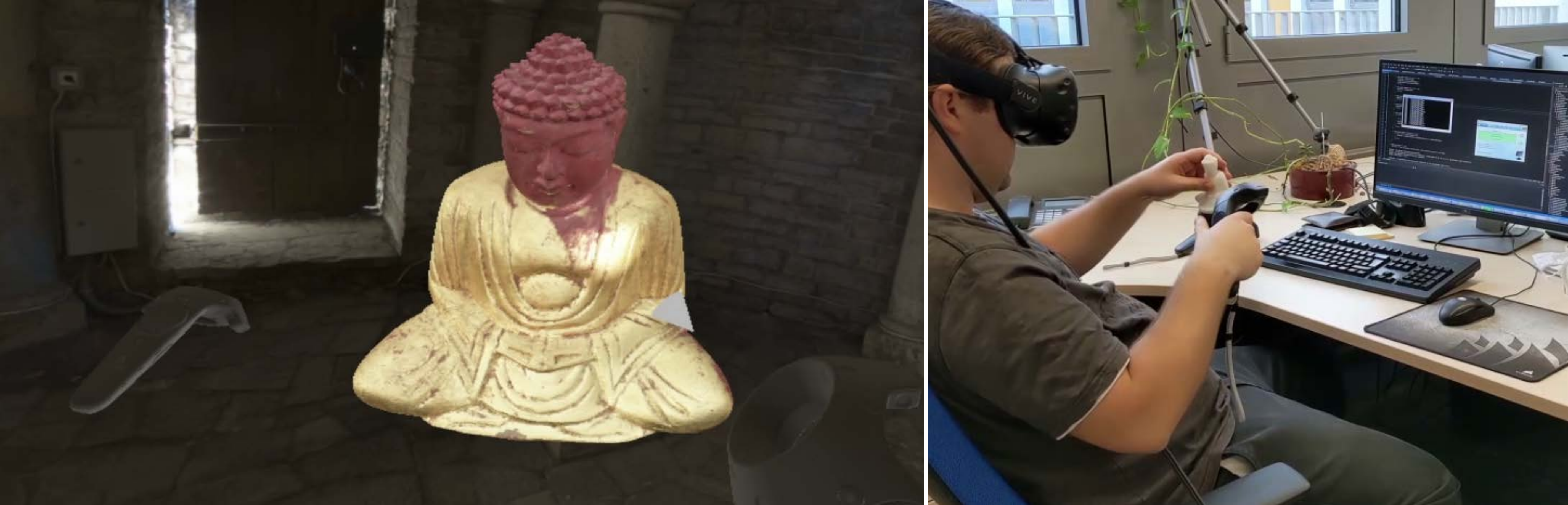}
	\caption{Illustration of our VR-Based Object Inspection system with haptic feedback.}
\end{teaserfigure}

\maketitle

\section{Introduction}
The potential of the rapidly emerging VR/AR technology lead to a new paradigm of presenting cultural heritage contents to the public and opens new opportunities regarding their analysis by expert and non-expert users.
Instead of an inspection and analysis solely based on depictions in static images, presenting digitized CH artifacts in terms of fully immersive and interactive experiences comes into reach.
The ability of creating an immersive experience of digitized artifacts is particularly relevant for the less stressful analysis/inspection of highly fragile objects, that may easily take damage during inspection, analysis or transport.
Furthermore, several artifacts are susceptible to certain environment conditions such as temperature, humidity or illumination conditions.
Hence, it is difficult or even impossible to include them in public exhibitions and complicates their use for scientific or educational purposes.
Instead, using accurately digitized virtual counterparts allows avoiding the use of the real artifact during inspection, analysis and transport and, additionally, opens new possibilities for simultaneous artifact inspection such as required for virtual museum experiences or collaborative analysis by experts situated at different physical locations.
However, the realization of such a virtual, optionally collaborative, object analysis and annotation depends on several crucial factors that have to be met to allow a pleasing experience while preserving as much of the aura of CH contents \cite{Benjamin:1935} as possible.
These challenges include the accurate object/scene digitization and representation, where both geometric object details as well as characteristic details of material appearance like reflectance behavior at scratches, engravings, etc. have to be captured, as well as intuitive interaction metaphors.
State-of-the-art reflectance representations based on memory-consuming bidirectional texture functions (BTFs)~\cite{Schwartz:2011,Schwartz:2014,Krumpen:2017,rainer2019neural,rainer2020neural} allow for an accurate virtual model of an object, but require powerful hardware for real-time rendering, whereas analytical reflectance models come at less computational effort and faster rendering.
This comes at the cost of often not sufficiently capturing relevant reflectance characteristics such as light exchange at fine structural details or local subsurface scattering.
Furthermore, the presentation of digitized contents on VR/AR devices imposes severe constraints regarding real-time rendering at framerates beyond 60 Hz on the visualization devices, instant/high-speed data exchange and interaction mechanisms to achieve a realistic impression for the joint analysis and interaction with the object or scene.
Despite this progress on representing and visualizing digitized objects, their visualization on screens or VR/AR glasses is impacted by the interaction devices (i.e. mouse, keyboard, controller, etc.) that limit the experience regarding the intuitive canonical interaction in terms of touching and rotating an object in your hands.
In this paper, we focus on enhancing the perceived aura of the digitized objects by augmenting the user experience with sensual information beyond purely sight-related data and providing additional haptic feedback regarding the object shape.
For this purpose, we present a practical tool for collaborative remote object analysis and annotation involving both the accurate depiction of the digitized artifacts at interactive framerates within VR devices and haptic feedback based on a 3D printed replicate to improve the inspection experience.
Using current VR devices, several experts located at possibly different physical places access and interact with high-quality virtual counterparts of real artifacts, where the reflectance data and the object geometry are stored on a central server and streamed to the remote clients as needed.
Our evaluation demonstrates that providing haptic feedback regarding the object shape enhances VR-based interaction, assessment and experience of the considered objects.
In summary, the main contributions of this work are:
\begin{itemize}
  \item We present a framework uniting the concepts of digital material appearance, VR technology and 3D printing that is designed for the needs of enhanced remote collaborative inspection of cultural heritage contents.
  \item Our interactive multi-client inspection and annotation tool allows multiple users at different physical places to collaboratively analyze objects based on various interaction metaphors for annotation and illuminating the object.
  \item We demonstrate the potential of our approach in the scope of a user study.
\end{itemize}

\section{Related Work}

\paragraph{Immersive Content Presentation}
Facilitating the access to CH contents by digital representations has been approached in various forms.
As the most obvious form given by still images or videos limits the user experience to pre-specified viewpoints and environment conditions (including the lighting conditions), the full aura of CH content is not preserved.
Instead, digitizing the 3D geometry based on laser scanning or structured light scanning and additionally capturing a digital representation of the reflectance behavior allows the manipulation of the environment and lighting conditions. Beyond interactive screen-based visualizations, a higher degree of immersion is achieved based on interactive 3D experiences of the CH objects/scenes. 
The range from real-world representations/environments to completely virtual representations/environments has been discussed under the notion of reality-virtuality continuum~\cite{milgram1994taxonomy}. 
For a detailed discussion of respective developments in the area of cultural heritage, we refer to the survey of by Bekele \etal~\cite{Bekele:2018}.

In VR settings, users are completely immersed into synthetic or pre-captured worlds that differ from the actual surrounding~\cite{carmigniani2011augmented,zhao2009survey}. Among the major applications for experiencing, exploring and manipulating CH contents are virtual museums and education. Augmented Virtuality aims at augmenting such completely virtual worlds with live real-world contents. In contrast, Augmented Reality (AR) enriches the actual real-world surrounding with synthetic contents as used for exhibition enhancement, education or exploration.

For all these approaches, interactions with the scene and its objects are realized based on the interplay between visual depictions and the control devices of VR/AR equipment.
However, these interfaces based on conventional VR/AR controllers, gamepads or gloves do not transport tactile feedback when interacting with objects.
Seminal work on tangible interfaces has been presented in terms of the Tooteko framework~\cite{dagnano2015tooteko} where facade reliefs were also presented to users as 3D printed objects lying on a table to allow haptic experience.
Touching different surface parts was coupled to respective audio feedback.
In contrast, our work extends VR-based 3D object inspection by additional haptic feedback based on a 3D-printed replica.

\paragraph{Physical Interactions in VR}
Early studies have demonstrated humans' ability to gauge movements of their own hands relative to their own bodies.
This \emph{proprioceptive sense}~\cite{Mine:1997} has been explored in terms of body-relative interaction types including the direct manipulation of objects as if these were in the user's hands, physical mnemonics (i.e. 3D body-relative widgets) and gesture commands.
Further work focused on providing a feedback to the user based on the shape of physical objects.
In such \emph{passive haptics} approaches~\cite{lindeman1999hand}, physical objects are aligned with objects in virtual environments which has been shown to enhance user experience of virtual environments~\cite{joyce2017passive}.
In contrast to providing a passive haptic feedback for static phenomena in the scene like ascending/descending stairs~\cite{Nagao:2018}, the interaction with dynamic objects relies on carefully tracking the respective real-world counterparts.
This can be achieved based on special gloves~\cite{Pierce:1999} as well as attaching tracking modules to physical objects~\cite{Franzluebbers:2018,hanus2019collaborative}.
Our approach also exploits passive haptics in terms of a 3D printed virtual counterpart of the object of interest, which is attached to a tracking module of typical VR equipment.
In contrast to previous work, our work focuses on both the accurate visual reproduction of objects within the virtual environments and the haptic feedback regarding the object's shapes.

\paragraph{Appearance Modeling}
Object appearance, i.e. the observed colors and textures, results from the complexity of light exchange at the surface depending on surface geometry, material reflectance characteristics and the surrounding illumination conditions.
As both the human visual system and acquisition devices are only capable of observing material appearance depending on the coupling of these three modalities, appearance modeling inevitably requires a decoupling of the respective modalities.
Respective acquisition strategies have been intensively discussed in literature~\cite{Weyrich:2009,Haindl:2013,weinmann-2016-EGCourse}.
While the geometric structure can be accurately recovered for a wide range of materials (i.e. diffuse to moderately specular surfaces) using structured light techniques or laser scanners~\cite{weinmann-2016-EGCourse}, the separation of reflectance and illumination properties remains challenging.
High-quality object digitization of smaller figurines is therefore mostly carried out in lab environments/using setups with controllable illumination such as gonioreflectometers~\cite{Holroyd:2010,filip2013brdf} or camera arrays~\cite{hawkins2001photometric,hawkins2005dual,Schwartz:2011,tunwattanapong2013acquiring,Schwartz:2014,kohler2013full,noll2013faithful,noll2015fully,havran2017lightdrum}.
Exploiting controllable illumination allows to accurately model reflectance in terms of parametric spatially varying bidirectional reflectance distribution (SVBRDF) models~\cite{nicodemus1977geometrical,Lensch:2003,Palma:2012,hawkins2001photometric,hawkins2005dual,holroyd2010coaxial,filip2013brdf,kohler2013full,noll2013faithful,noll2015fully,santos2014cultlab3d,nam2018practical,ladefoged2019spatially} or data-driven bidirectional texture functions~\cite{Dana:1997,Schwartz:2011,Schwartz:2014,havran2017lightdrum,Krumpen:2017} that both represent spatially varying surface reflectance depending on view and illumination conditions.
SVBRDFs are parameterized over the exact 2D surface of the object geometry and usually the local surface reflectance behavior is represented in terms of an analytical function.
This makes them particularly suitable for specular materials, where the highlights may be lost with a tabulated representation.
However, as the accuracy of the scanned geometry is still limited regarding fine structural details, the light exchange may not be accurately reproduced in such cases. 
In contrast, BTFs are parameterized over an approximate object surface. Hence, subtle effects of light exchange including interreflections, self-shadowing and self-occlusions induced by finer structures not preserved in the reconstruction as well as local subsurface scattering are directly stored in terms of a data-driven image-based representation.
The latter results in significantly increasing computational and memory requirements in comparison to parametric representations.
Whereas standard BTFs are parameterized over 2D surface domains which suffers from distortions or seam artifacts, the 3D parameterization of OctreeBTFs~\cite{Krumpen:2017} avoids distortions and significantly reduces the number of seam artifacts.
While our framework is not restricted to a particular representation, we employ the data-intensive OctreeBTF representation~\cite{Krumpen:2017} to demonstrate that even such expensive representations can be used for interactive object inspection and annotation.
In addition to a purely visual (in our case VR-based) experience, we provide additional haptic feedback to enhance the user experience.
Further digitization methods in the context of cultural heritage are based on reflectance transformation imaging (RTI)~\cite{Malzbender:2001,Mudge:2006,Mudge:2008,Earl:2010,Ponchio:2018,giachetti2018novel,dulecha2020neural}.
However, these techniques usually capture the lighting-dependent reflectance of an object only from a single view.
Hence, RTI techniques are not adequate for in-hand object inspection scenarios, where the view-dependent object appearance has to be taken into account as well.

\paragraph{Rendering and Streaming of Appearance Data}
Collaborative object inspection by users at different phyiscal locations relies on efficient model representations, fast data streaming as well as rendering.
Storing raw BTF measurements comes at huge memory requirements of several tens to hundreds of gigabytes.
These can be compressed based on the combination of fitting a mixture of several SVBRDFs to the BTF data and residual apparent BRDFs (ABRDFs)~\cite{wu2011sparse}.
The latter are required to take non-local effects of the light transport in the material into account as otherwise the quality of the reproduced reflectance behavior may be severely impacted.
As an alternative, several compression techniques rely on interpreting the BTF as a tensor for which a low-rank approximation can be found via factorization.
Respective techniques include Full Matrix Factorization (FMF) \cite{koudelka2003acquisition}, Decorrelated-FMF (DFMF)~\cite{Schwartz:2013} and K-SVD based compression~\cite{ruiters2009btf}, that are capable of preserving details of material appearance and allowing real-time rendering.
However, the K-SVD approach outperforms the FMF-based approaches by a compression factor of about 3 to 4 for comparable quality.
Furthermore, investigations on perceptually-inspired BTF compression leveraged the observation that lower downsampled levels of details sufficiently approximate some of the factorized data, thereby reducing the memory requirements on the GPU wich led to compression factors of about 500.
Further BTF compression schemes rely on the use of neural networks~\cite{rainer2019neural,rainer2020neural}, where the compression and decompression are achieved based on an autoencoder approach, or on statistical approaches~\cite{haindl2007extreme}, that allow extreme compression at the cost of loosing the capability to accurately reproduce fine characteristics of reflectance behavior, as well as vector-quantization-based approaches~\cite{havran2010bidirectional} or the use of  multivariate radial basis functions~\cite{tsai2010modeling}, that both reach compression rates similar to FMF-based compression.
Aside from BTF compression to allow for real-time rendering, scenarios where BTFs have to be streamed over the internet also should allow for a progressive refinement of the rendering. 
This can be achieved by leveraging the fact that the used DFMF-compression orders the BTF data with respect to its importance to stream a BTF over the internet and progressively refine the rendering when more data arrives~\cite{Schwartz:2013}.
Further work has shown that the spatial data parts of a 2D-parameterized BTF can be split into smaller tiles that can be streamed to the GPU independent of each other~\cite{Schwartz:2013_2}, which enables only loading the BTF data for the object regions that are currently visible on the screen.
In turn, this enables rendering scenes containing of multiple BTFs in real-time, which would otherwise exceed the amount of VRAM available if each BTF would have to be loaded completely without streaming.
In our system, we combine these ideas~\cite{Schwartz:2013,Schwartz:2013_2} to allow for a progressive refinement of the rendering of OctreeBTFs~\cite{Krumpen:2017}, while streaming the most important data, \ie the data that contributes most to the final result first.

\section{Methodology}

\begin{figure*}[t]
    \centering
    \includegraphics[width=\textwidth]{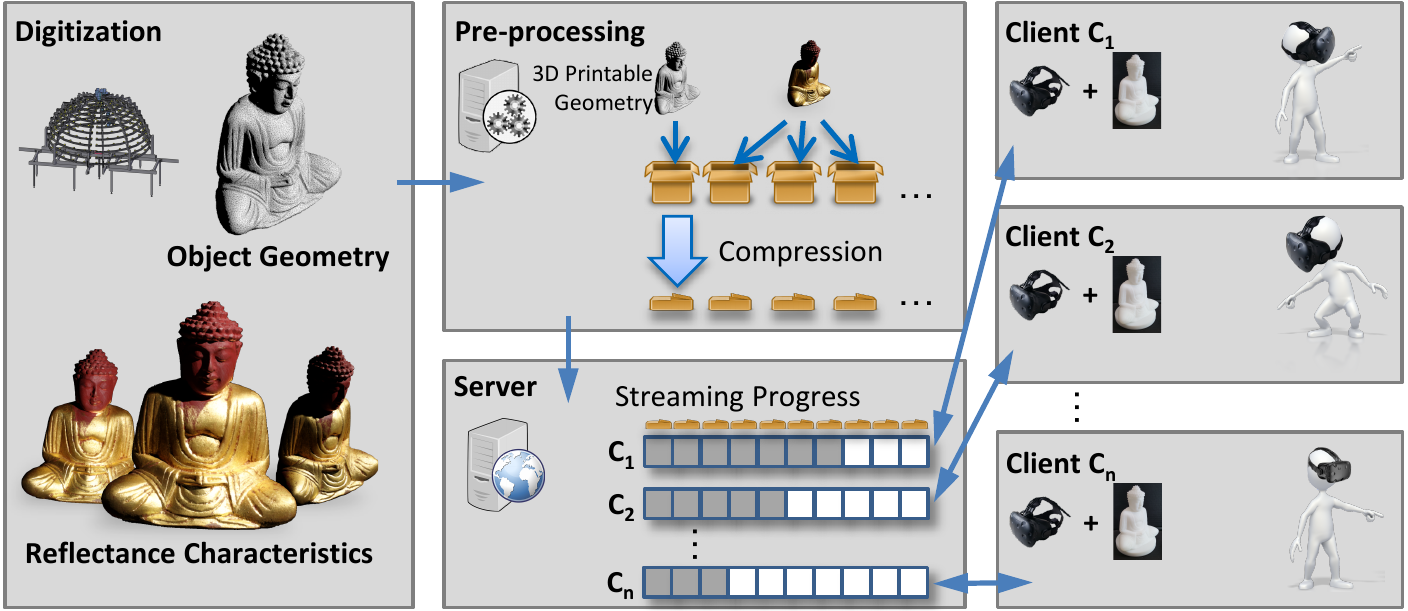}
    \caption{Overview of our approach towards sharing tangible cultural heritage experiences. The initial digitization outputs the object geometry, the reflectance characteristics in terms of an OctreeBTF \cite{Krumpen:2017}. Then, a pre-processing step (upper center) computes a version of the geometry suitable for 3D printing and converts the BTF data into a streamable version which is loaded by a server (bottom center). The server then streams the compressed OctreeBTF to remote clients that can interactively inspect the object in VR, enhanced by haptic feedback provided by the 3D printed replica of the object.}
    \label{fig:streaming_pipeline}
\end{figure*}
Our approach for the joint immersive, tangible experience of CH contents by multiple users at possibly different physical locations relies on a a client-server model as illustrated in Figure~\ref{fig:streaming_pipeline}.
Inputs in terms of the geometric model as well as the reflectance characteristics (for the purpose of an accurate object representation, we use the OctreeBTF representation~\cite{Krumpen:2017}) of the considered models have to be obtained from a prior digitization process or a model repository.
In an initial step, we perform a pre-processing of these data by computing a 3D printable version of the 3D shape and transform the OctreeBTF into a more compressed representation that allows for progressive streaming over a network.
The resulting object representation (i.e. in terms of geometry and reflectance) is stored on the server and streamed to each remote client which renders the object on a HMD and allows for interactive inspection enhanced with haptic feedback from a 3D printed replica.
In the following sections, we discuss the involved components in more detail.
As we address the generation of highly immersive experiences of cultural heritage objects, the objects need to be accurately captured so that characteristic details that determine the object appearance remain preserved in the digital model.
State-of-the-art capture techniques~\cite{Weyrich:2009,Haindl:2013,weinmann-2016-EGCourse} rely on an initial 3D scanning of the object geometry using laser scanners or structured light systems.
Subsequently, spatial variations of local surface appearance are captured under different viewing and lighting conditions.
Depending on the surface material characteristics, the dependence of the local reflectance behavior $f(\mathbf{x},\boldsymbol{\omega}_v,\boldsymbol{\omega}_l)$ at a surface point $\mathbf{x}$, the viewing direction $\boldsymbol{\omega}_v$ and the lighting direction $\boldsymbol{\omega}_l$ can be represented using Spatially Varying Bidirectional Reflectance Distribution Functions (SVBRDFs)~\cite{nicodemus1977geometrical} or Bidirectional Texture Functions (BTFs)~\cite{Dana:1997}.
Our framework is not restricted to one particular representation.
SVBRDFs as captured in several other investigations~\cite{Lensch:2003,Palma:2012,hawkins2001photometric,hawkins2005dual,holroyd2010coaxial,kohler2013full,noll2013faithful,noll2015fully,santos2014cultlab3d,nam2018practical,ladefoged2019spatially} could also be integrated.
However, to demonstrate the efficiency of our approach, we conduct our experiments with the computationally more demanding OctreeBTF representation~\cite{Krumpen:2017} since it delivers the best visual quality of the representations mentioned above.
The OctreeBTF representation relies on parameterizing the reflectance representation in a volumetric manner instead of a 2D parameterization over the surface which ameliorates seam artifacts and texture distortions. 
We use datasets captured with a camera array~\cite{Schwartz:2011,Schwartz:2013} for which surface geometry has been reconstructed using a structured light system~\cite{Weinmann:2011} and Octree-BTF representations have been provided by the authors of~\cite{Krumpen:2017}.
In addition to the reflectance characteristics, the capturing process also outputs the detailed geometry of the object.

\subsection{Pre-processing for Streaming}
\label{ssec:streaming}

In a first step, the captured 3D geometry is converted to a version that is suited for 3D printing. 
In particular, if the capturing process does not output a closed mesh, possibly occurring holes in the geometry have to be filled.
Afterwards, the model is scaled to a size that fits on the tracker-object (see Figure~\ref{fig:printed_objects}), that is part of the VR system and used to track the printed object pose.
Optionally, if the object does not have a flat bottom surface or the surface does not cover the extends of the tracker, a socket can be added in order to mount the object on the tracker, as can be seen in Figure~\ref{fig:printed_objects}.
Furthermore, we have to consider the fact that the typical capturing process outputs the OctreeBTF in a way which is well-suited for real-time rendering, but does not allow for (progressive) streaming over a network due to data size and data layout.
Hence, before we store the OctreeBTF on the server, it has to be further compressed and converted to a representation that both allows for progressive streaming and real-time rendering.
Therefore, the BTF representation is split into parts that solely depend on the spatial position $\mathbf{x}$ and the light- and view-directions $(\boldsymbol{\omega}_v,\boldsymbol{\omega}_l)$ respectively.
These parts are further divided into \textit{chunks} that represent different levels of the octree corresponding to levels of detail of the reconstruction.
The resulting data chunks as well as the geometry are compressed by using Zstandard~\cite{Collet:2017:zstd} and ordered in a scheme that allows the client to start rendering when the first few chunks have been transmitted, and to progressively refine the rendering as more chunks arrive.
For a more detail on the BTF compression, how the data is divided into the spatial and angular parts and how the chunks are ordered, we refer to the supplemental material.

\subsection{Server Component}
Initially, the server loads the compressed BTF data chunks and the geometry and then waits for clients to connect. 
Each new client first receives general information about the object, followed by the geometry and the BTF data chunks in the order discussed in the supplemental material.
The server keeps track of the chunks transmitted to each client, allowing multiple clients to connect at different times.
Whenever a client places an annotation on the object, the annotation is forwarded to all other clients, and is also stored on the server, so that clients connecting at a later time also receive all annotations already present.

\subsection{Remote Client Application}
At the client side, we use an off-the-shelf VR-system consisting of a HMD, two controllers and an additional tracker (see Figure \ref{fig:printed_objects} right), that can be attached to an object to include said object into the virtual world. 
To allow insights on how the haptics-enriched VR experience improves on a purely screen-based experience, we also created a Non-VR version of the application (in the following denoted as the "`2D version"'), where the object and the light are controlled by moving the mouse.

\paragraph{Tracking}
The VR-System is set up in such way that we can use a simple tracker for the object, which is held in one hand, and a controller for the light and menu interactions held in the other hand.
The tracker features a standard 3/4 inch mount used for cameras that we use to fix the printed object on the tracker, which has the benefit of allowing to quickly exchange the objects.
We utilize the trackpad of the controller to control the light intensity and the object scale. 
Each rendered frame, the tracker, the controller and the HMD send their positions and orientations to the application, which then adapts the scene to match the positions in the real world.

\paragraph{Rendering and Streaming}
When the client has connected to the server, a placeholder geometry is rendered until the real geometry has been received and decompressed.
In general, decompression is performed in separate thread on the CPU, and the uncompressed data is transferred to the GPU in an asynchronous manner, allowing for seamless updates of geometry and reflectance information without stuttering, which is specifically important when rendering on VR-headsets.
When a chunk has been uploaded to the GPU, the render-thread is signaled to update the rendering according to the now loaded data.
Due to the iterative streaming of BTF data, the user can already view the object and place annotations on it, while the rendering is refined as more data arrives.

\subsection{3D Printing for Haptic Feedback}
\label{ssec:3Dprinting}
\begin{figure}[t]
    \centering
     \includegraphics[width = 0.8\linewidth, height = 0.5\textheight, keepaspectratio]{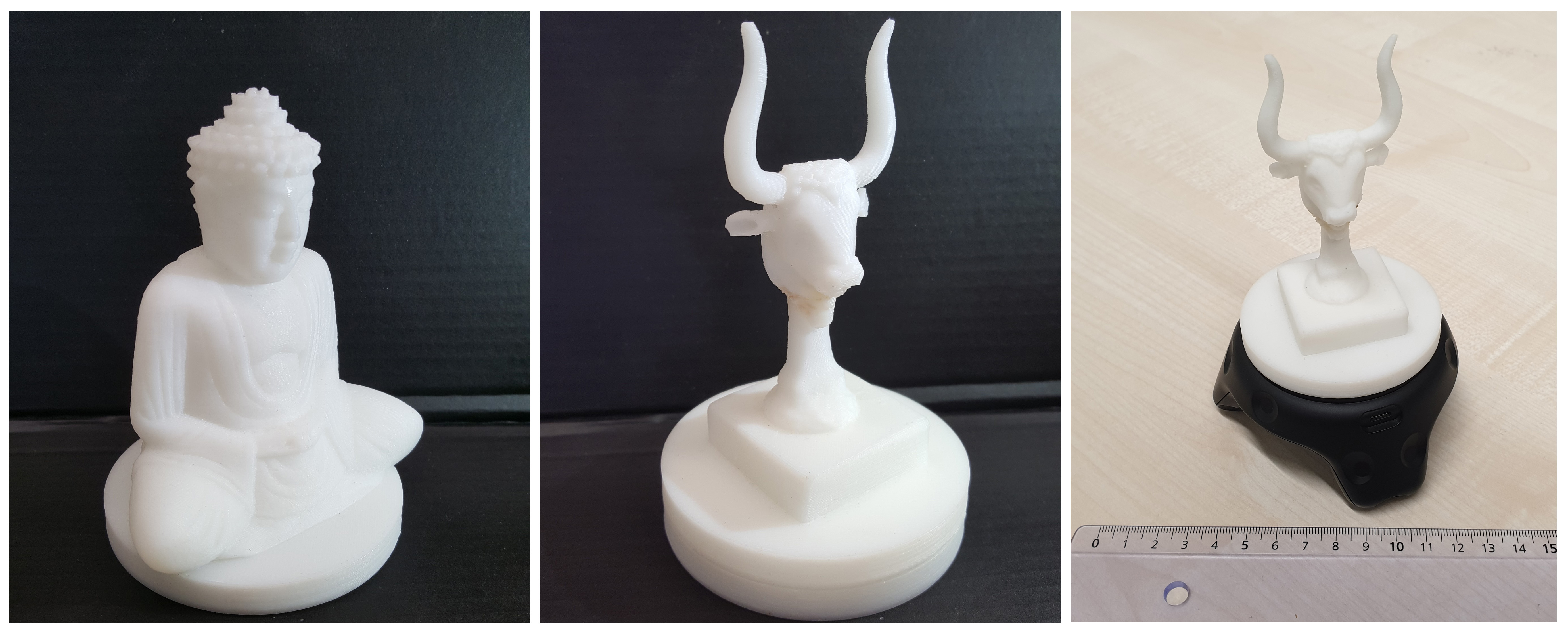}
		\caption{Left and middle: Printed objects, right: Object mounted on the tracker.}
		\label{fig:printed_objects}
\end{figure}

To provide the user with haptic feedback, we chose to print the model using an off-the-shelf 3D-printer.
This printer uses the Fused filament fabrication (FFF, also known as fused deposition modeling, FDM) technique, where a thermoplastic material is heated to the point where it melts, and deposited by a moving printhead in a per-layer fashion.  
The advantage of that technique is, that over the last years, printers have become more and more affordable and easy to use, while the quality of the printed objects increased.
As we want to attach the objects to the tracker, we add a baseplate to the object with a hole for a screw. Printed objects are shown in Figure~\ref{fig:printed_objects}.
Furthermore, the objects have to be scaled down to fit the tracker, since a large object would occlude the tracker too much, thereby decreasing the tracking accuracy.

\section{Interaction Tools for Remote Inspection}

The major challenge for the design and development of interaction metaphors are given by the specification of desirable interaction possibilities with objects as well as their intuitive realization based on hardware equipment.
To allow an immersive object visualization, we leverage the potential of current VR devices and extend the interaction possibilities by adding haptic object experience.
In the following, we provide details regarding the used control mechanisms for handling the object of interest as well as changing illumination conditions as well as adding annotations.
The respective controls can easily be adapted for right-handed and left-handed persons (in the following, we will only describe the typical setting for right-handed persons).

\paragraph{Controls for Object Handling}
As mentioned in Section~\ref{ssec:3Dprinting}, we create a haptic feedback during object exploration in VR by 3D printing the respective object geometry and attaching it on a VR tracking device (we use the HTC Vive tracker, to which the printed object is attached).
To allow a comparison regarding how haptic feedback from the printed object enhances the user experience, we also implemented a second mode where the digital object is (virtually) attached to one of the standard controllers, i.e. this interaction is solely determined by how the VR controller is moved and does not provide any haptic feedback.
The application is implemented in a way so that we can switch between these two modes at any time directly within the application.
In addition, we allow the object to be scaled in the virtual environment based on the trackpad of the controller assigned to the right hand.

\paragraph{Light Controls}
An important aspect for object exploration is given by the respective illumination conditions.
Besides showing how an object will look like within a certain environment, which we address by environment lighting~\cite{debevec2006_course}, our viewer also supports the direct lighting of an object via a user-controlled light source during its inspection.
When using the direct lighting mode, the controller in the right hand is used to control the main light source that can be either a point light, or a spotlight to simulate a flashlight as commonly used to inspect objects.
The intensity of the light source can be changed via the controller. 
In case of environment lighting, we use a spherical environment map to define the incoming light from every direction.
However, as correct image-based lighting is not possible for BTFs in real-time, we approximate the environment map by a series of eight directional light sources.

\begin{figure}[t]
    \centering
     \includegraphics[width = 0.8\linewidth, height = 0.5\textheight, keepaspectratio]{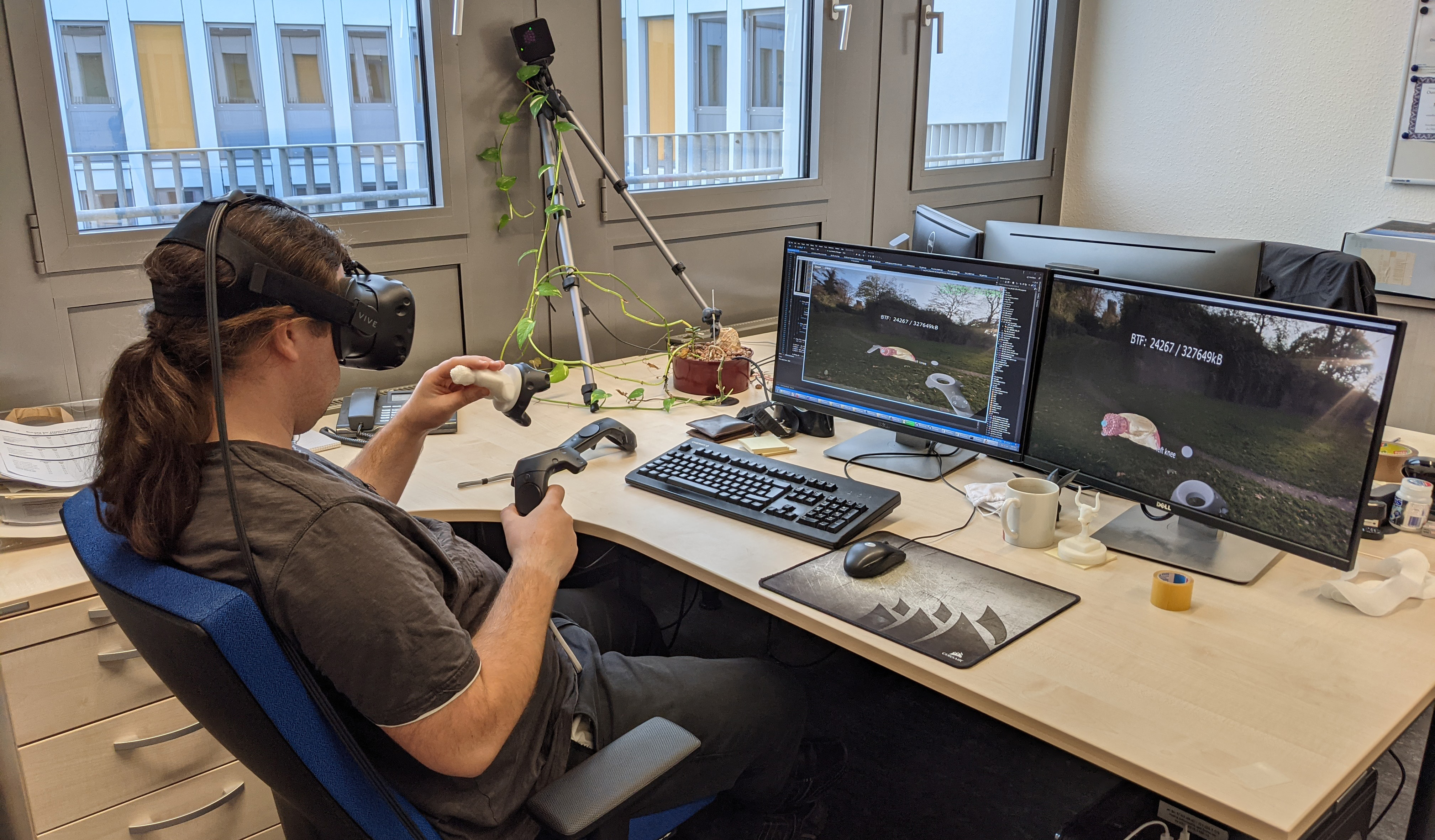}
		\caption{View on a person operating the system.}
		\label{fig:interact}
\end{figure}

\paragraph{Annotation Tools}
During artifact inspection, users often need to annotate respective object parts.
For this purpose, our client application allows users to place annotations on the object by moving the second controller (i.e. the one where the object is not attached to) over a spot on the object, and pressing a button on the controller.
Users can thereby annotate the object either by placing a simple marker highlighting a spot to direct other (possibly remotely connected) users to inspect this in more detail, or, if more information is needed, a short and descriptive text can be added to the annotation using a virtual keyboard.
Annotated points on the object are displayed as small spheres and the color of the spheres reflects the type of annotation, i.e. whether its purpose is the highlighting of that particular location or region to other users or whether additional text information is stored at the respective location.
The latter texts are only visible to a user if she moves the controller over or next to the sphere to avoid continuously occluding the object by the text.
All added annotations are sent to the server, which, in turn, forwards them to other connected clients.
The display of annotations can be activated or deactivated at any time.\\
As a further functionality, we allow the user to draw on the object by continuosly placing points on the object surface while a certain controller button is pressed.
This could be used for marking larger features on the object.
\section{Experimental Results}

To demonstrate the potential of our approach, we provide a perceptual evaluation obtained in terms of a user study as well as evaluations regarding visual quality and performance. In addition, we compare our enhanced VR system to a purely 2D screen-based application with the same features.

\subsection{User Study}
\label{sec:study}
\begin{figure*}
    \centering
		\includegraphics[width=\linewidth, keepaspectratio]{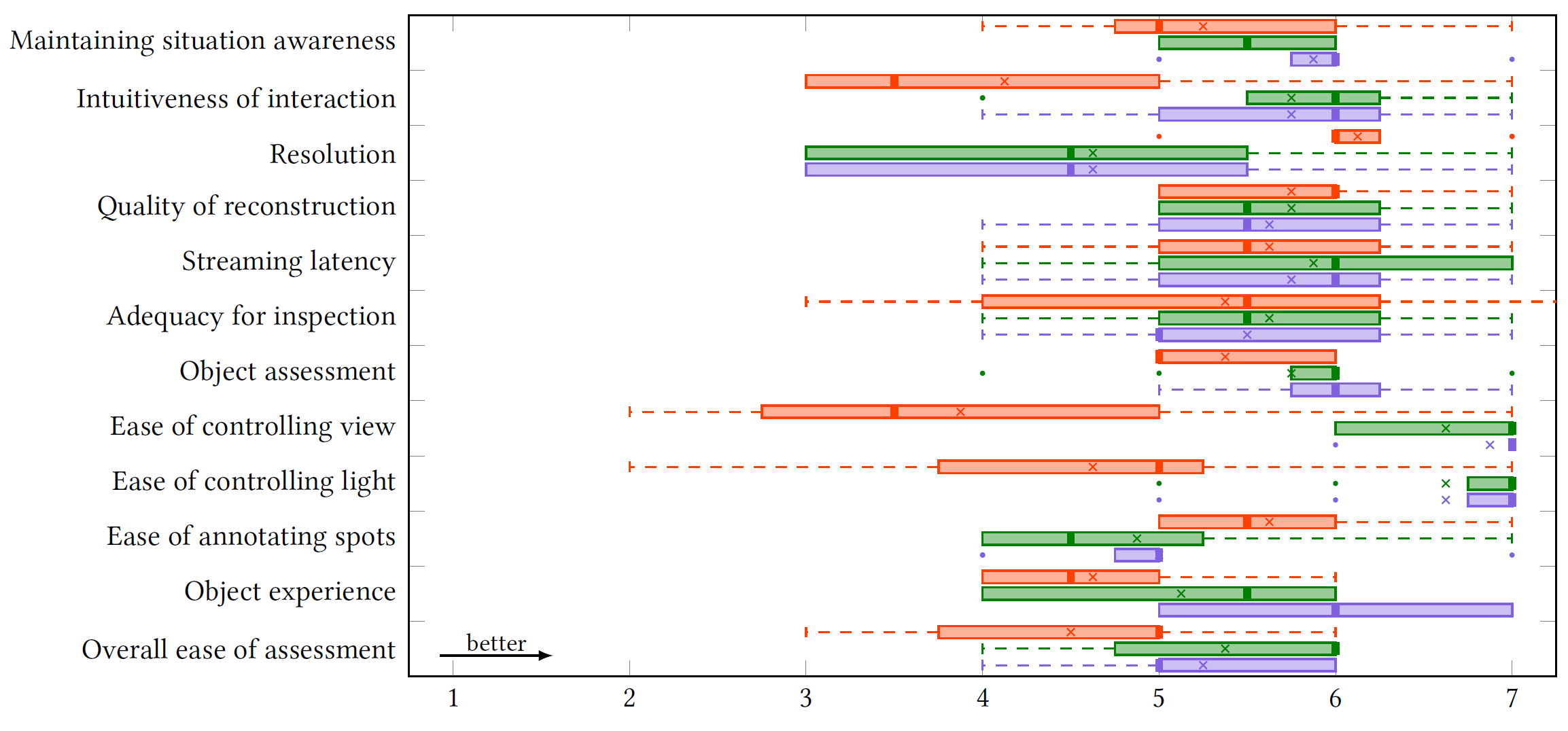}  
    \caption{Statistical results of our user study. Statistics regarding ratings obtained for mode $M_1$ are colored in red, the ones for mode $M_2$ in green and the ones for mode $M_3$ in blue. The illustration includes lower and upper fence (dashed line), interquartile range (colored bar), median (marked with a vertical bar),  outliers (marked with $\bullet$) as well as the average value (marked with $\times$). For most aspects, the VR-based modes $M_2$ and $M_3$ outperform the 2D application ($M_1$) with keyboard-mouse controls. The additional use of haptic feedback of the object geometry further improves the quality of object experience and assessment as well as view control and leads to the best overall ratings.}
	\label{fig:study_results}
\end{figure*}

To verify whether the combination of VR-based object visualization as well as haptic feedback actually enriches the experience of inspecting cultural artifacts, we performed a user study where the participants had to inspect an object and perform certain tasks, such as annotating spots or describing materials or reflectance behavior of the respectively considered object.
All participants were naive to the goals of the experiment, provided informed consent, reported normal or corrected-to normal visual and hearing acuity. 
During the experiment, the participants had to perform the given tasks in three different modes:
\begin{itemize}
  \item In the first mode $M_1$, the 2D application was used where the user can manipulate the object displayed on a standard screen based on mouse/keyboard interactions.
	\item In the second mode $M_2$, users could interact with our novel VR-based application without haptic feedback (i.e. the virtual model was attached to the second controller of the VR-system instead).
	\item In the third mode $M_3$, users were able to interact with the objects based on our VR-application and the printed model to allow for an additional haptic feedback.
\end{itemize}
To avoid biases, we varied the order of these modes for individual participants, i.e. the participants started with different modes.
For each condition, the participants were asked to provide ratings on a 7-point Likert scale regarding aspects such as visual quality, adequacy for inspection, quality of object experience and ease of interacting with the object, performing annotations or controlling view and light conditions.
The results shown in figure \ref{fig:study_results} demonstrate that the approaches $M_2$ and $M_3$ outperform the 2D application $M_1$ in terms of the ease of inspecting the object and controlling light conditions and, hence, result in a better overall experience and intuitiveness regarding the interactions, which may be a result of the higher intuitiveness of hand-based interactions with objects. The latter represent a kind of canonical approach for us humans to access real 3D objects in our daily lifes.
In comparison to the depiction of objects on a screen and keyboard/mouse based interactions ($M_1$), users particularly acknowledged the easiness of controlling the viewpoint based on VR-based systems $M_2$ and $M_3$.
Furthermore, users seemed to benefit from additionally using a printed replica ($M_3$) regarding the view control.
Mode $M_3$ additionally also received the best ratings regarding the quality of object assessment and object experience which were likely to also result in the best overall ease of assessment.
However, the resolution of the 2D application $M_1$ was rated higher than for the VR modes $M_2$ and $M_3$, which is a result of the limited resolution of the used HMD. 
Finally, we expect the lower scores regarding the annotations observed for Modes $M_2$ and $M_3$ in comparison to $M_1$ to be the result of the absence of a standard keyboard for entering text. Here, the use of a virtual keyboard, where a laser-pointer is used to target each letter individually would have to be exchanged by a respective speech recognition system to rapidly allow adequate text annotations. However, as this aspect does not touch the main insights to be gained in the scope of this investigation -- that the use of haptic feedback within a VR-based object exploration improves the object experience -- we leave it for future work.

\subsection{Intuitiveness of Interaction}
\begin{figure*}
    \centering
     \includegraphics[width=\linewidth,keepaspectratio]{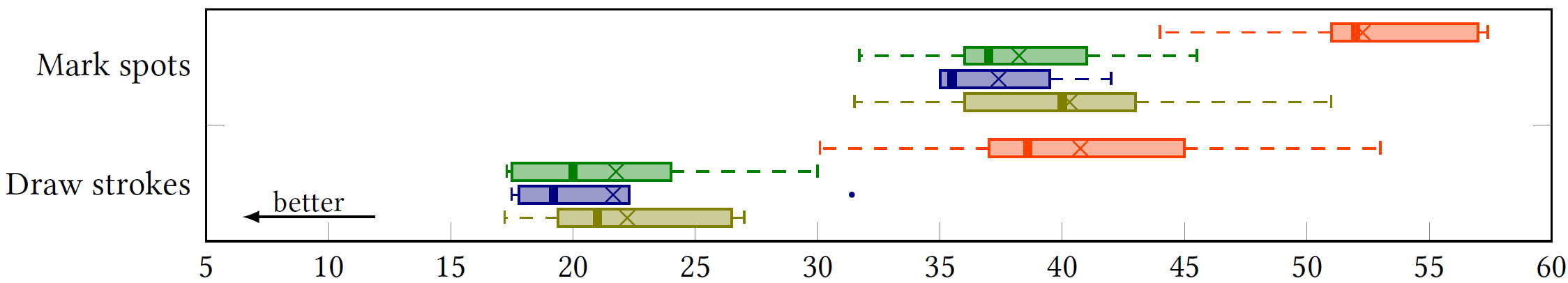}
    \caption{Comparison of interaction times for marking a fixed series of spots on an object as well as drawing strokes along characteristic object features. Interaction times for 2D application ($M_1$) are shown in red, times for $M_2$ are given in green, values for the VR mode with haptic feedback ($M_3$) are given in red the new mode $M_4$ is printed in yellow.}
    \label{fig:study_results2}
\end{figure*}

In a further experiment, we focused on the analysis whether the improved intuitiveness of interaction reported in our previous study can also be seen in the times required for users to perform certain tasks, that involve an interaction with the object such as rotating it.
To further evaluate if providing the exact geometry for haptic feedback benefits the interaction, we also provided a simple geometry in terms of a 3D printed cube which was mounted on the tracker (denoted as $M_4$).
In this experiment, the participants were first asked to highlight a fixed series of spots on the object by clicking on the object and placing an annotation.
To separate the impact of the object interaction from the textual annotation, we only focused on the times to mark the respective surface points and  discarded the entering of text information for this experiment.
Each participant performed this task in all the modes $M_1$, $M_2$, $M_3$ described in Section~\ref{sec:study} as well as the new mode $M_4$, starting with alternating modes to eliminate biases.
We constructed several series of points the users had to annotate, and alternated the series between the modes and participants to eliminate learning effects.
The series of spots were defined in such a way, that the object has to be rotated in order to annotate the next spot, and when done, the participants were asked to rotate the object back to its original position.
In addition, we let the users draw strokes along characteristic object features (neck, face and legs of the Buddha figurine) in all four modes.
The respectively required interaction times are displayed in Figure~\ref{fig:study_results2} and indicate that providing any form of immersive presentation is beneficial for navigation when the object has to be rotated in order to perform a certain task.
Further providing haptic feedback regarding the object shape gives an additional small benefit, that can be seen in the variance of the interaction times.

\subsection{Visual Quality}
\begin{figure*}
		\centering
		\includegraphics[width=\linewidth, keepaspectratio]{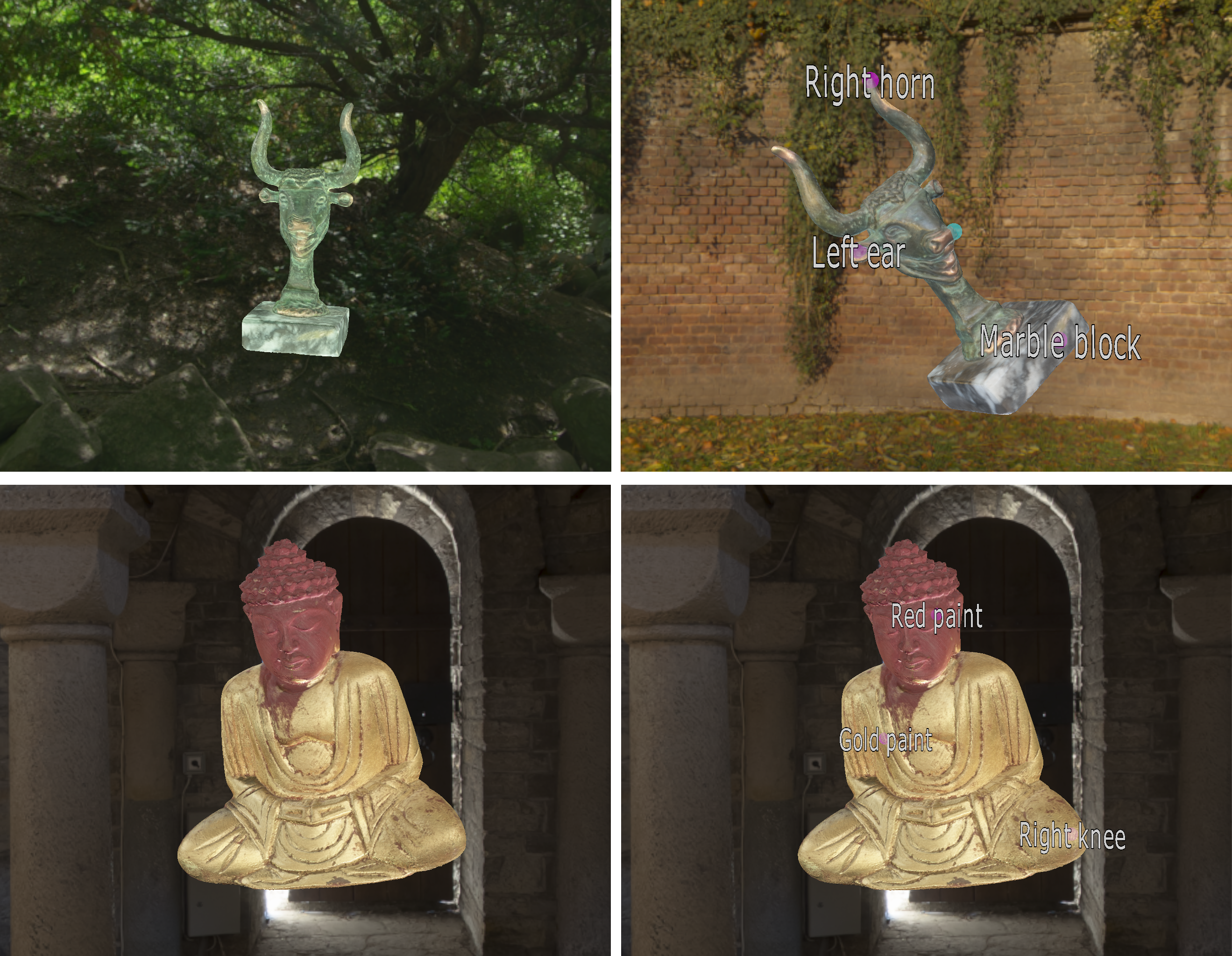}  
		\caption{Screenshots of the 2D application  where two objects are lit by different environments with the user-controlled light being disabled. In the right images some object parts have been annotated by the user. Please note that the textsize has been increased while capturing the screenshots for better readability. In the real application, the texts are much smaller.}
		\label{fig:screen_label}
\end{figure*}

\begin{figure*}
    \centering
    \includegraphics[width=\linewidth]{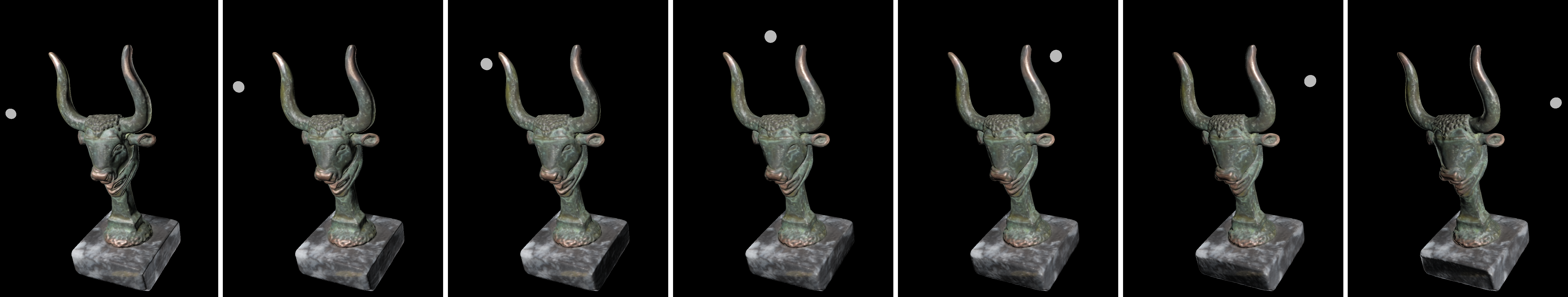} 
    \caption{A series of screenshots of the 2D application where the environment lighting is disabled and the user-controlled light is moved on an arc above the object, showing how the reflectance changes according to the light position.}
	\label{fig:lighting}
\end{figure*}

As we considered a reflectance representation in terms of OctreeBTFs~\cite{Krumpen:2017}, there are no seams and distortions that affect the visual quality in comparison to the use of conventional BTFs parameterized over the 2D object surface.
However, we had to turn off the spatial filtering of OctreeBTFs in the VR mode to meet the performance requirements for interactive VR-based visualization.
Despite this, the limited OctreeBTF resolution only becomes visible when the object is held very close to the HMD.
In Figure \ref{fig:screen_label}, we show a set of screenshots from the 2D version of the application, where different objects are placed in different environment lighting and some surface points have additionally been annotated by a user.
Furthermore, the dependence of object appearance under varying illumination conditions (in this example, the light source is moved) is shown in Figure \ref{fig:lighting}.

\subsection{Performance analysis}
To analyze the performance of our approach, we measured the average frametimes over a period of about 30 seconds during a typical interaction scenario as seen in Figure \ref{fig:interact}, both using the 2D application and the VR version.
When rendering an OctreeBTF, or BTFs in general, the main bottleneck is the amount of data that has to be read from the VRAM each frame, which means that rendering is mainly bandwidth limited.
The amount of data mainly depends on the number of screen pixels covered by the object, which determines how much of the spatial BTF data has to be fetched from the VRAM, and the number of light sources used to approximate the illumination as we need to sample the angular data for each light source.
To get insights regarding how much the additional lights used to represent the environment (in our experiments, we approximated environments using eight light sources) affect the performance, we performed the measurement once without considering environment lighting and once with the respective environment.
When interacting with the objects, we additionally took typical distances for close-range inspection into account to provide a realistic scenario, which means that a large fraction of the screen pixels of the virtual camera of the 2D application or, in the VR case, the HMD are covered.
As shown in Table \ref{table:framerates}, frametimes are as expected higher when rendering in VR, even exceeding 28ms, leading to 35 frames per second, i.e. a slight stuttering of the rendering, due to the increased resolution and the fact that we have to render the scene twice to allow the stereo impression.
However, interestingly, this aspect was not perceived to be disturbing by the participants during the user study (see Section~\ref{sec:study}).
We further noticed that after the BTF is fully streamed, and the last octree level has been uploaded to the GPU, the frametimes increase significantly.
This is due to the fact that at this point there is an increase in the amount of data the GPU has to fetch from the VRAM for each pixel.
In addition, we observed that the frametimes decrease again when the object is really close to the camera.
This happens when one voxel of the octree covers more than one screen pixel, which means that the overall number of visible voxels decreases as well as the amount of data the GPU needs to fetch from the VRAM.
Then, the data can be cached between shader invocations that process the same voxel.
But for this effect to occur, the object must be held really close to the camera, especially in VR, and the spatial resolution of the BTF becomes visible.
\begin{table}
\caption{Average frametimes {[}ms{]} and frames per second {[}Hz{]} for both the 2D application and the VR applications. The values were measured during a typical interaction scenario.}
\label{table:framerates}
\begin{tabular}{lcc}
\toprule
										& \textbf{2D}                   & \textbf{VR} \\  
\midrule
With environment    & 18.18 / 55                    & 28.50 / 35   \\
Without environment & 14.28 / 70                    & 18.18 / 55   \\
\bottomrule
\end{tabular}
\end{table}
\subsection{Limitations}
Despite the benefits of allowing an improved, \emph{tangible} remote object experience with particularly better capabilities regarding interactive object inspection and object assessment, our approach also faces limitations.
One major limitation is the resolution of current VR-HMDs, which is an issue that will hopefully be resolved in the near future by the next generations of HMDs.
On the other end, higher display resolutions also demand for higher resolution of the used OctreeBTF, which, in turn, requires for more powerful GPUs with more video memory, and better compression techniques for real-time streaming.
Also, the used 3D printing process has some limiting factors. The use of an FDM-printer, which prints objects layer by layer, causes these layers to be perceivable when touching the object and probably impact the haptic feedback regarding fine geometrical features of the object.
A solution would be to use a SLA-printer, which can produce much thinner layers and thus can better preserve small structures of the object. 
However, we only expect a small impact on the perception in the scenarios shown in the scope of our evaluation, as the respective object features were easy to perceive, and that the overall tendency of an improved tangible object experience will only be slightly improved.
Furthermore, most 3D printers are limited to one material which is usually plastic, but also wood like materials are possible.
Some printers have multiple extruders, which allow for multi-material prints, but these materials are all of the same kind, e.g. one cannot mix plastic and wood-materials.
We hope that in the future, these issues can be resolved by more advanced printing technology, and, hence, do not consider them to be permanent limitations of our overall approach towards creating tangible object experiences.

\section{Conclusions}

We have presented an approach towards tangible cultural heritage experiences.
Our method allows interactive and collaborative VR-based object inspection and annotation based on high-quality 3D models with accurate reflectance characteristics, while additionally providing haptic feedback regarding the object shape features based on a 3D printed replica.
As demonstrated by our user study, the additional haptic feedback enriches VR-based interaction, assessment and experience of the considered objects, which, in turn, indicates the potential of our approach.
While the navigation speed is on par with alternative VR-based interaction methods, providing haptic feedback regarding the object shape enhances the overall experience.
In future work, we would also see the possibility to circumvent the use of a hardware tracking device, e.g. based on printing patterns used for tracking into the geometry.

\bibliographystyle{ACM-Reference-Format}
\bibliography{submission}

\appendix
\newpage

\section{BTF compression}
BTF measurements can be stored in terms of huge matrices, columns contain the reflectance values for a particular surface point $\mathbf{x}$ under various view-light configurations $(\omega_l, \omega_v)$.
Depending on the resolution of the measurement setup, even when stored with half-precision floating point values, this matrix can exceed hundreds of gigabytes, and thus, needs to be compressed in order to meet the demands for real-time rendering and streaming.
For this, the Decorrelated Full Matrix Factorization (DFMF)~\cite{mueller-2009-dissertation}, has been demonstrated to achieve good compression ratios, while preserving visual quality. 
In a first step, the BTF color values are converted into the YUV colorspace, which separates brightness from color information. 
Subsequently, the color channels are decorrelated by taking the logarithm of the Y-channel, and dividing the U- and V-channel by the Y-channel. 
Then, for each of these matrices, a SVD $A = \mathbf{U} \times \mathbf{\Sigma} \times \mathbf{V}^T$ is computed to separate spatial information stored in $\mathbf{V}$ from the light- and view-dependent information in $\mathbf{U}$.
More precisely, a column $u_c(\omega_l, \omega_v)$ of $\mathbf{U}$ stores the \emph{Eigen-ABRDF}, whereas a row $v_c(\mathbf{x})$ of $\mathbf{V}$ holds the \emph{Eigen-Texture} of the BTF component $c$.
As the SVD orders the rows $\mathbf{V}$ and the columns of $\mathbf{U}$ with respect to their importance, we can compress the BTF by keeping the first $k$ rows of $\mathbf{V}$ and the first $k$ columns of $\mathbf{U}$.
\[
 \rho_{BTF}(\mathbf{x},\boldsymbol{\omega}_l,\boldsymbol{\omega}_v) \approx \sum_{0 \leq c < k}{u_c(\boldsymbol{\omega}_l,\boldsymbol{\omega}_v) \cdot \boldsymbol{\sigma}_c \cdot v_c(\mathbf{x})}
\]
The resulting matrices $\widetilde{\mathbf{U}}$, $\widetilde{\mathbf{\Sigma}}$ and $\widetilde{\mathbf{V}}$ for each color channel are then stored, where $\widetilde{\mathbf{\Sigma}}$ and $\widetilde{\mathbf{V}}$ are pre-multiplied for simplicity.
Using this compression, the BTF data size is reduced to a few gigabytes, which fit into the memory of current GPUs. Furthermore, the use of the YUV colorspace allows to compress the color channels U and V even further by keeping less components, as most reflectance information is about changes in brightness and the human eye is more sensitive to these than to changes in color.
In our experiments, we observed the use of eight BTF components for the U- and V-channel to be a good compromise between data size and visual quality when rendering.
While for the Y-channel a maximum of 100 components is sufficient for nearly all objects and for objects with less details, one can also take less components.
When rendering, four components are grouped together so the number of components $k$ is chosen to be a multiple of four, an RGBA-array-texture with $k / 4$ layers can be used for rendering. 
Applying the above, we end up with $k_Y = 72$ and $k_{UV} = 8$.
This algorithm can be applied to both conventional, texture-based BTFs as well as OctreeBTFs. The difference is, that for OctreeBTFs, the compression is done for the highest octree-level $d$, and the compressed BTF data for the lower levels is computed as a post-processing step by averaging the BTF data over all child nodes in a top-down manner, starting with level $d-1$ down to $d_{min}$ as described by Krumpen \etal~\cite{Krumpen:2017}.
\section{BTF data layout}
Even when compressed, transmitting a BTF to a client over the network still takes some time, which makes it feasible to split the BTF into smaller chunks of data and transmit these chunks independently to a remote client.
This enables the client to start rendering the BTF with a lower quality as soon as the first chunks are received and to refine the rendering as more data-chunks arrive. 
An approach for streaming BTFs over the internet has been presented yb Schwartz et al.~\cite{Schwartz:2013}, however, since the data-layout of the spatial matrices $\mathbf{\widetilde{\Sigma}} \times \mathbf{\widetilde{V}}$ of the OctreeBTF differs from a conventional BTF, this approach cannot be applied to stream OctreesBTFs. 
The four-dimensional angular data from $\mathbf{\widetilde{U}}$ is stored as a nested parabolic map converted to an RGBA-array-texture for both BTF representations and the layers of these textures can be streamed independently.
The difference when streaming lies in the handling of spatial data:
For conventional, texture-based BTFs, the spatial data can be stored as an RGBA-array texture for each color channel, where each layer of a texture stores the BTF components $c_i \ldots c_{i+3}$ for all surface points $\mathbf{X}$. The individual texture-layers can then be streamed independent from each other, which means that the client can start rendering as soon as the first texture-layer for all three color channels is received. 
For OctreeBTFs however, the data is stored per voxel, where one voxel corresponds to one point $\mathbf{x}$, i.e. first all components $c_1^0 \ldots c_k^0$ for the first voxel are stored, then components $c_1^1 \ldots c_k^1$ for the second voxel, etc.
This data layout leads to a better cache-efficiency on the GPU during rendering, as described by Krumpen et al.~\cite{Krumpen:2017}, but has the disadvantage that it is no longer possible (at least not without significant, costly re-arrangement of the data) to stream one component for all voxels as one data-chunk. 
Thus, we have to treat all components for all voxels for one octree levels as one data-chunk, and the client can only start rendering if the whole chunk (again for all color channels) has been received.
Finally, rendering the OctreeBTF needs information about the octree itself and the normal and tangent vectors for each voxel, leading of an additional chunk of voxel-data per octree level.
We construct the OctreeBTF similar to Krumpen et al.~\cite{Krumpen:2017}, where the first octree levels are forced to be complete, thus we have $l = d_{max} - d_{min}$ octree levels. 
Considering the number of texture-layers for the angular data for all three color channels, we end up with
\begin{itemize}
	\item $l$ chunks of voxel data, storing the octree structure.
	\item $l \times 3$ chunks of spatial BTF data, which cannot be split into components.
	\item $k_Y/4 + 2\times k_{UV}/4$ chunks of angular BTF data, \ie texture-layers, which are independent from the octree structure.
\end{itemize}
All chunks are further compressed using the Zstandard library~\cite{Collet:2017:zstd} and are stored in the main memory of the server, together with the also compressed geometry.
\section{OctreeBTF streaming}
\label{ssec:streaming}
For streaming, we need to determine an order in which the chunks are transmitted, such that the client can start rendering as early as possible.
First, the geometry is transmitted, which is followed by the voxel and spatial data for the octree level $d_{min}$ and the first four components of angular data (\ie texture-layer) for each color channel.
The remaining angular texture layers are distributed evenly among the remaining octree-levels.
The biggest chunk is the last spatial chunk for the Y-channel due to the large number of voxels with $k_Y$ components for each. 
The complete algorithm is provided in Algorithm \ref{algo:load_order}.
\begin{algorithm}
	\caption{Build the load order for BTF data chunks}
	\label{algo:load_order}
	\begin{algorithmic}[5]
		\STATE Add voxel data for level $j_0$
		\STATE Add spatial data for level $j_0$ and channels YUV
		\STATE Add first angular-layer for channels YUV
		\STATE layers\_per\_lvl\_Y = max(layers\_total\_Y / $l$, 1)
		\STATE layers\_per\_lvl\_UV = max(layers\_total\_UV / $l$, 1)
		\STATE layers\_added\_Y = 1
		\STATE layers\_added\_UV = 1
		\FOR{$i=1$ to $l$}
			\STATE Add voxel data for level $j_i$
			\STATE Add spatial data for level $j_i$ and channel Y
			\FOR {$g = 1$ to layers\_per\_lvl\_Y}
				\IF {layers\_added\_Y < layers\_total\_Y}
					\STATE Add angular-layer[layers\_added\_Y]
				\ENDIF
				\STATE {layers\_added\_Y+=1}
			\ENDFOR
			
			\STATE Add spatial data for level $j_i$ and channel U
			\STATE Add spatial data for level $j_i$ and channel V
			\FOR {$g = 1$ to layers\_per\_lvl\_UV}
				\IF {layers\_added\_UV < layers\_total\_UV}
					\STATE Add angular-layer[layers\_added\_UV]
					\STATE Add angular-layer[layers\_added\_UV]
				\ENDIF
				\STATE layers\_added\_UV+=1
			\ENDFOR
		\ENDFOR

	\end{algorithmic}
\end{algorithm}

\end{document}